# Measurement of the Background Activities of a $^{100}$Mo-enriched Powder Sample for an AMoRE Crystal Material by using Fourteen High-Purity Germanium Detectors

S. Y. Park[a, b], K. I. Hahn[c, d], W. G. Kang[b], V. Kazalov[e], G. W. Kim[b], Y. D. Kim[b, f, g], E. K. Lee[b], M. H. Lee[b, f, *] and D. S. Leonard[b, **]

[a]Department of Physics, Ewha Womans University, Seoul, 03760, Korea
[b]Center for Underground Physics, Institute for Basic Science (IBS), Daejeon, 34126, Korea
[c]Department of Science Education, Ewha Womans University, Seoul, 03760, Korea
[d]Center for Exotic Nuclear Studies, Institute for Basic Science (IBS), Daejeon, 34126, Korea
[e]Baksan Neutrino Observatory, Institute for Nuclear Research of the Russian Academy of Science, Kabardino-Balkaria, 361609, Russia
[f]IBS School, University of Science and Technology (UST), Daejeon, 34113, Korea
[g]Department of Physics and Astronomy, Sejong University, Seoul 05006, Korea

*Corresponding author (M. H. Lee): Tel.: +82 42 878 8518; fax: +82 42 878 8509.
E-mail address: mhlee@ibs.re.kr
**Corresponding author (D. S. Leonard): Tel.: +82 42 878 8534; fax: +82 42 878 8509.
E-mail address: dleonard@ibs.re.kr

**Abstract**

The Advanced Molybdenum-based Rare process Experiment in its second phase (AMoRE-II) will search for neutrinoless double-beta (0νββ) decay of $^{100}$Mo in 200 kg of molybdate crystals. To achieve the zero-background level in the energy range of the double-beta decay Q-value of $^{100}$Mo, the radioactive contamination levels in AMoRE crystals should be low. $^{100\text{Enr}}$MoO$_3$ powder, which is enriched in the $^{100}$Mo isotope, is used to grow the AMoRE crystals. A shielded array of fourteen high-purity germanium detectors with 70% relative efficiency each was used for the measurement of background activities in a sample of 9.6-kg powder. The detector system named CAGe located at the Yangyang underground laboratory was designed for measuring low levels of radioactivity from natural radioisotopes or cosmogenic nuclides such as $^{228}$Ac, $^{228}$Th, $^{226}$Ra, $^{88}$Y, and $^{40}$K. The activities of $^{228}$Ac and $^{228}$Th in the powder sample were 0.88 ± 0.12 mBq/kg and 0.669 ± 0.087 mBq/kg, respectively. The activity of $^{226}$Ra was measured to be 1.50 ± 0.23 mBq/kg. The activity of $^{88}$Y was 0.101 ± 0.016 mBq/kg. The activity of $^{40}$K was found as 36.0 ± 4.1 mBq/kg.

## 1. Introduction

Searches for neutrinoless double-beta (0νββ) decay can potentially determine if the neutrino is a Dirac or Majorana type particle and may provide information about the mass and mass hierarchy of neutrinos. Observation of the 0νββ decay is only possible in a handful of isotopes where the background-producing single β decays are energetically forbidden. Because of its high Q-value and reasonably high natural abundance (9.7%), $^{100}$Mo is a notable candidate. If the 0νββ decay occurs, two electrons emitted from the 0νββ decay of $^{100}$Mo will carry the full energy of the double-beta decay Q-value of 3,034.4 keV [1]. However, the 0νββ decay is known to be rare, with the half-life > $1.1 \times 10^{24}$ year [2].

A flagship experiment of the Center for Underground Physics (CUP) at the Institute for Basic Science (IBS) is the Advanced Molybdenum-based Rare process Experiment (AMoRE) [3, 4]. The second phase of AMoRE (AMoRE-II) will be prepared to search for the 0νββ decay of $^{100}$Mo in 200 kg of molybdate crystals. To observe two electrons from the 0νββ decay, the AMoRE-II requires a background level < $10^{-4}$ count/keV/kg/year in the region of interest (ROI) around the Q-value [5].

The energy resolution of the AMoRE detectors is about 10 keV in full width at half maximum (FWHM) [6] and the ROI around the Q-value is from 3,024.4 keV to 3,044.4 keV. The activities of all radioisotopes having Q-values around or over 3,034.4 keV that contribute background signals in the ROI must be known or constrained. The decay of $^{208}$Tl, which occurs in equilibrium with $^{228}$Th from the $^{232}$Th decay chain, is one of the most dangerous background sources for AMoRE. This decay emits two gammas with energies of 583.2 keV and 2,614.5 keV, having total energy of 3,197.7 keV [7]. Partial collection of this summed energy may result in background signals in the ROI. Also, $^{214}$Bi in equilibrium with $^{226}$Ra from the $^{238}$U decay chain is a source of backgrounds since its β$^-$ decay Q-value is 3,270 keV and its α decay Q-value is 5,621 keV [8, 9]. Decays of cosmogenic nuclides can also produce background signals. The isotope $^{88}$Y is made by cosmogenic activation and can produce significant background signals in the ROI [10]. The β$^+$ decay Q-value of $^{88}$Y is 3,622.6 keV [11].

Sufficiently high-activity decays with low Q-values can create backgrounds in the ROI through random coincidence with other low energy depositions of similar or different origins. The naturally occurring and ubiquitous β$^-$ decay of $^{40}$K has a Q-value of 1,310.9 keV [12]. In coincidence with other activities, including two-neutrino double-beta (2νββ) decay of $^{100}$Mo, this can also produce significant backgrounds if it is present at high enough levels.

The internal background levels of the AMoRE crystals were measured by acquiring the activities of $^{216}$Po in equilibrium with $^{228}$Th from the $^{232}$Th decay chain and $^{214}$Po in equilibrium with $^{226}$Ra from the $^{238}$U decay chain [13-15]. Background-producing activities in materials and components of the AMoRE detector system were measured [16, 17]. Decays of radioisotopes were simulated to determine, based on these measurements, which parts of the internal AMoRE detector system contribute the most to the experiment's backgrounds [18]. In Ref. 18, internal radioactive sources in the AMoRE crystals were found to produce potentially dominant contributions to the background signals in the ROI.

The levels of radioactive contaminants in the raw materials, including the $^{100\text{Enr}}$MoO$_3$ powders used to grow the AMoRE crystals, must be understood to control the background. The AMoRE crystals are grown from MoO$_3$ powder after purification by using physical and chemical methods [19]. Measuring radioactivity impurity levels of the initial and purified powders facilitates a quality-control before the crystal growth as well as providing input for the purification requirements and feedback on the purification process.

Two shielded 100% relative efficiency HPGe detectors named as CC1 and CC2 are operating at the Yangyang underground laboratory (Y2L) [16, 17, 20]. Activities of unstable isotopes in a powder sample of 1.6 kg were initially measured using the CC1 detector [21]. The activities of $^{228}$Ac and $^{228}$Th were both < 1.0 mBq/kg at a 90% confidence level (C.L.). The activity of $^{226}$Ra was measured to be 5.1 ± 0.4 (stat) ± 2.2 (syst) mBq/kg, with a notably high systematic error relative to the obtained value. The $^{40}$K activity was found to be < 16.4 mBq/kg at 90% C.L. Upper limits were obtained for $^{228}$Ac, $^{228}$Th, and $^{40}$K.

A shielded array named CAGe comprising fourteen HPGe detectors with 70% relative efficiency each and having a total relative efficiency of 980% was constructed in the Y2L [22-25]. This system was designed for rare physics process searches as well as trace-radioactivity measurement in potential detector materials [16, 20, 24]. The CAGe has nearly ten times the total efficiency of either CC1 or CC2 and has a large sample volume. Better measurement sensitivities are generally expected from the CAGe than from CC1 or CC2 detectors. The other advantage of the CAGe is the ability it provides to tag coincident events, such as a coincidence between the 583.2 keV and 2,614.5 keV gammas from the $^{208}$Tl decay. This can be potentially advantageous for background discrimination in a number of scenarios. We collected data for 1,803.3 hrs of CAGe operation with a 9.6-kg sample of powder.

## 2. Experimental Setup

The construction and installation of the CAGe detector system, electronics of the CAGe, and its performance are explained in Ref. 25. The CAGe has two cryostats facing each other. The upper and lower cryostats with their connected detectors are referred to as the 'top array' and 'bottom array', respectively. The top array, its dewar, and the upper part of the shielding are movable to optimize the vertical space between the two arrays for different-size samples. The distance between the opposing endcaps of the two arrays was set to 54 mm for the installation of the powder sample.

About 15 kg of the $^{100\text{Enr}}$MoO$_3$ powder for AMoRE was delivered from JSC Isotope in each shipment. A 15-kg powder shipment contains 10 kg of molybdenum enriched to 95.73% $^{100}$Mo by mass, as certificated by the supplier. The $^{100\text{Enr}}$MoO$_3$ powder was packed in twenty-one separate plastic zipper bags. Fourteen bags were selected, considering their sizes and how they would fit in the CAGe sample volume. Six bags containing a total of 3.8 kg of the $^{100\text{Enr}}$MoO$_3$ powder were installed in the space between the top array and the bottom array. They were supported on a 5 mm thick acrylic plate, with three acrylic legs supporting the plate from the lower cryostat as indicated in Fig. 1(a). Eight bags containing a total of 5.8 kg of $^{100\text{Enr}}$MoO$_3$ powder surrounded the CAGe. A total of 9.6 kg of the powder sample was placed in the CAGe as shown in Figs. 1(b) and 1(c).

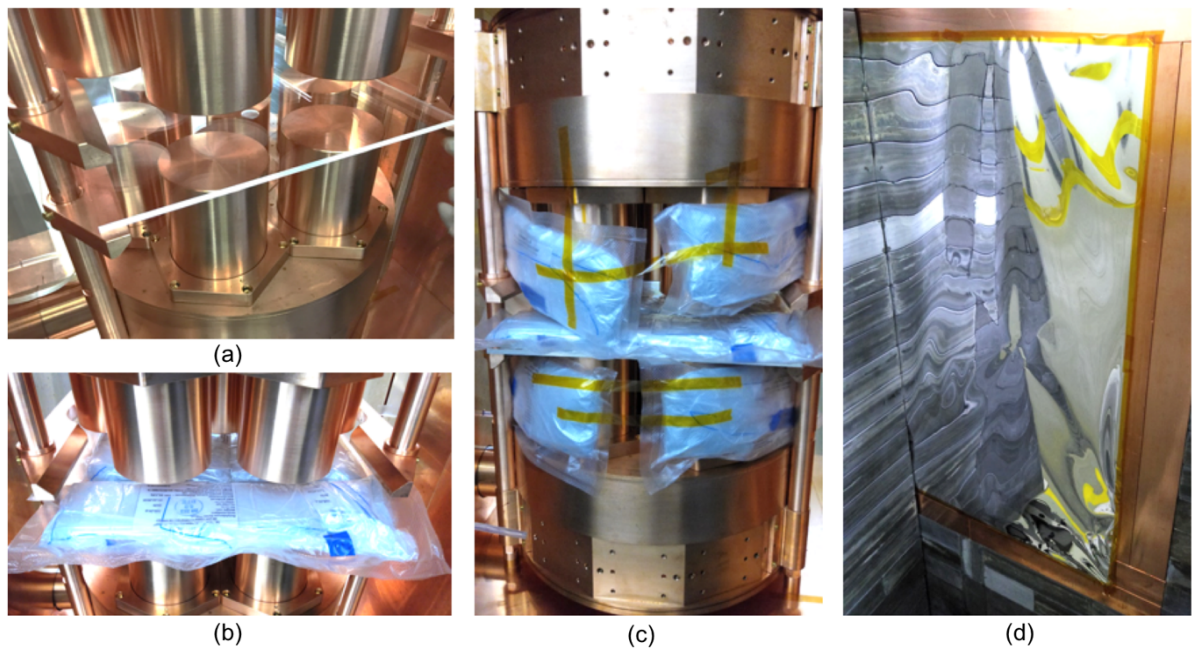

Fig. 1. (a) An acrylic plate and three legs for supporting the powder sample, (b) mid-powder-sample installation, (c) the powder sample surrounding the CAGe, and (d) a Vikuiti sheet sealing one side of the detection volume.

Radon gas entering the detection volume from the air in the laboratory can produce elevated and unstable background signals from the $^{226}$Ra sub-chain, resulting in reduced sensitivity and worse systematic errors. To reduce this effect, the detection volume was continuously flushed with nitrogen gas from the boil-off of a liquid nitrogen bottle. 3M-brand Vikuiti sheets shown in Fig. 1(d) sealed the detection volume, keeping a high concentration of nitrogen gas in the volume. A radon reduction system (RRS) supplied air to the laboratory with concentrations of $^{222}$Rn significantly reduced relative to the tunnel environment [26].

The RRS was turned off during the first (P1) and the third (P3) data taking periods but was operating during the second period (P2). Fig. 2 shows the radon activity in the laboratory measured by a RAD7 detector [27] during the entire measurement period. Observed activities were $0 – 5$ Bq/m$^3$ and $20 – 50$ Bq/m$^3$ for the RRS on and RRS off periods, respectively. The total live time was 922.7 hrs, 523.1 hrs, and 357.7 hrs for P1, P2, and P3, respectively.

Background data without the powder sample were obtained to subtract signal contributions from contaminants in and around the CAGe detector system. Two consecutive background data points were taken with the same experimental conditions used for the powder data, including the presence of the acrylic sample supporter, the Vikuiti sealing, and the nitrogen gas flushing. The RRS was turned off during the first (B1) background data taking period but was operating during the second (B2). The total live time was 663.1 hrs and 919.4 hrs for B1 and B2, respectively.

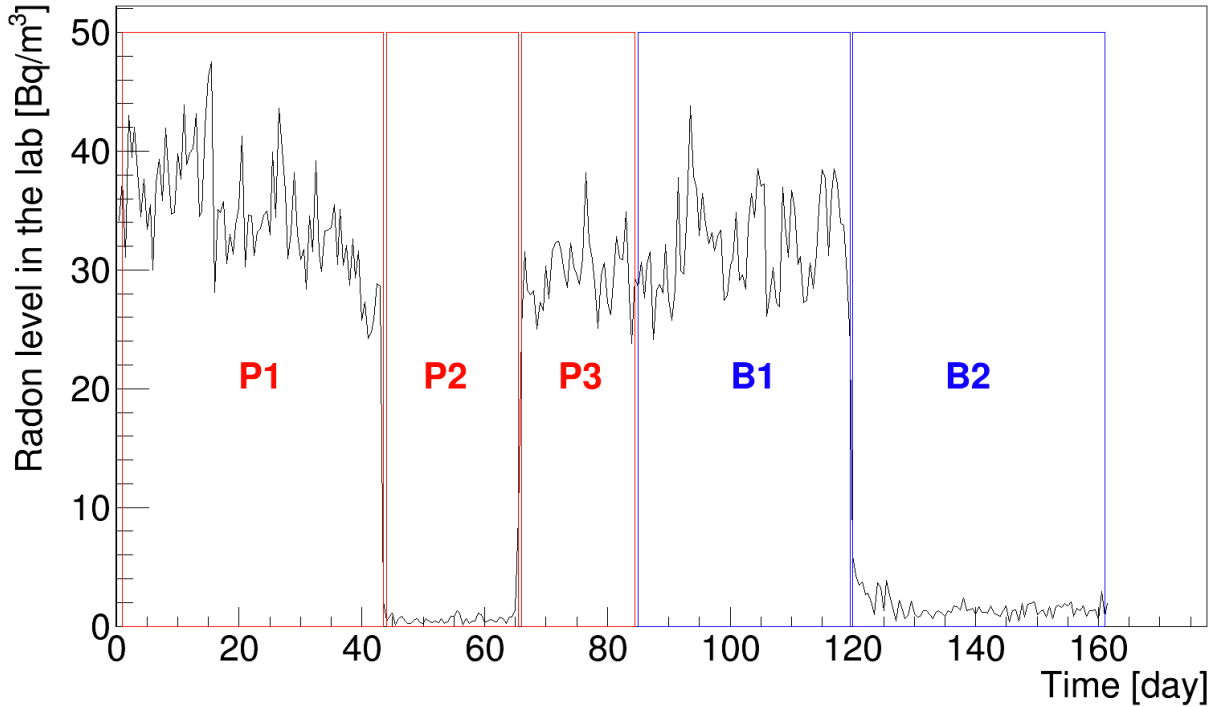

Fig. 2. Radon level during the powder and the background data taking. P1, P2, and P3 are for data taking periods of the powder sample in red boxes, B1 and B2 are for those of background in blue ones.

### 3. Energy Calibration and Resolution

For data period P1, the energy was calibrated with the most probable values from fits of the 295.2 keV peak from $^{214}$Pb, the 352.0 keV, 609.3 keV, and 1,764.5 keV peaks from $^{214}$Bi, and the 1,460.8 keV peak from $^{40}$K. For the counting intervals P2 and P3, four peaks of 352.0 keV, 609.3 keV, 1,764.5 keV, and 1,460.8 keV were fitted for the energy calibration. For energy calibrations of B1 and B2, the 511 keV peak from annihilation, the 1,332.5 keV peak from $^{60}$Co, and the 1,460.8 keV peak from $^{40}$K were used.

After calibrating the energy, we checked the energy resolution of each HPGe detector element using the 609.3 keV and 1,460.8 keV peaks. The corresponding FWHM values for data periods P1, P2, and P3 are listed in Table I. The FWHMs are typically smaller than 2.5 keV and 2.9 keV at the peaks of 609.3 keV and 1,460.8 keV, respectively [28, 29]. However, resolutions for detectors 2 and 4 in the bottom array were substantially higher than those for other detectors. To allow for fitting a single combined spectrum with a single resolution, data from detectors 2 and 4 were excluded in the powder activity analysis.

**TABLE I**

FWHM resolutions for each detector module. Values are shown for the three data P1, P2, and P3.

| Array | Detector number | FWHM of 609.3 keV [keV] | | | Detector number | FWHM of 1,460.8 keV [keV] | | |
|---|---|---|---|---|---|---|---|---|
| | | P1 | P2 | P3 | | P1 | P2 | P3 |
| Bottom | 1 | 1.60 | 2.18 | 1.65 | 1 | 2.35 | 2.02 | 2.37 |
| | 2 | 1.72 | 2.42 | 3.13 | 2 | 2.57 | 3.37 | 2.99 |
| | 3 | 2.00 | 2.11 | 1.67 | 3 | 2.60 | 2.27 | 2.12 |
| | 4 | 3.71 | 4.21 | 3.48 | 4 | 4.08 | 3.69 | 3.58 |
| | 5 | 1.69 | 1.52 | 1.92 | 5 | 2.54 | 2.71 | 2.45 |
| | 6 | 1.53 | 1.56 | 1.67 | 6 | 2.71 | 2.31 | 1.92 |
| | 7 | 1.86 | 2.05 | 1.62 | 7 | 2.27 | 2.19 | 2.23 |
| Top | 8 | 2.10 | 2.08 | 2.41 | 8 | 2.78 | 2.19 | 2.11 |
| | 9 | 1.55 | 1.88 | 1.73 | 9 | 2.30 | 2.39 | 2.01 |
| | 10 | 1.54 | 1.59 | 1.52 | 10 | 2.10 | 1.89 | 2.02 |
| | 11 | 1.68 | 1.76 | 1.83 | 11 | 2.48 | 2.38 | 2.48 |
| | 12 | 1.71 | 1.73 | 1.81 | 12 | 2.34 | 2.20 | 2.17 |
| | 13 | 1.59 | 1.80 | 1.70 | 13 | 2.07 | 2.05 | 2.17 |
| | 14 | 1.65 | 1.51 | 1.71 | 14 | 2.44 | 2.03 | 2.26 |

To confirm the uncertainty of the energy calibrations, energy spectra from twelve detectors were summed into one combined spectrum in each data set of P1, P2, P3, B1, and B2. The 1,332.5 keV peaks in each data set were fitted to Gaussian distribution functions added to a parameterization of the continuous backgrounds caused by Compton scattering. The obtained central values for the Gaussian distributions were compared to the known precise value of 1,332.492(4) keV from $^{60}$Co in Ref. 30. The peak positions were higher than the known value by 0.007 keV, 0.104 keV, and 0.003 keV for P1, P2, and P3, respectively, and by -0.002 keV and 0.008 keV for B1 and B2, respectively.

### 4. Influence of the Radon Reduction System

Background signals for most of the $^{226}$Ra sub-chain, except for the weak gamma peak from $^{226}$Ra itself, can be contributed by $^{222}$Rn decays in the detection volume, and thus the background rates can fluctuate based on the radon levels. Fig. 3 shows the combined spectra from 40 keV to 3000 keV for P1 (RRS off), P2 (RRS on), and P3 (RRS off). The continuous background level below 500 keV is lower in P2 than in either P1 or P3. On the other hand, the continuum rates of P1 and P3 are similar to each other.

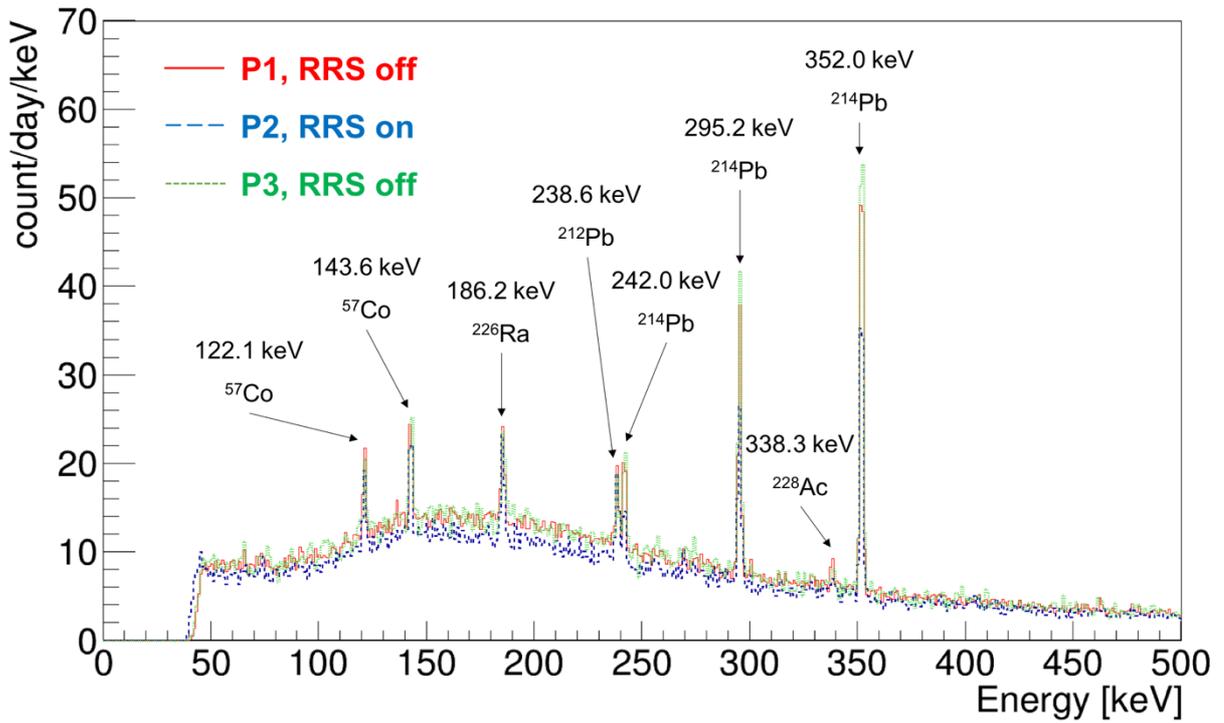
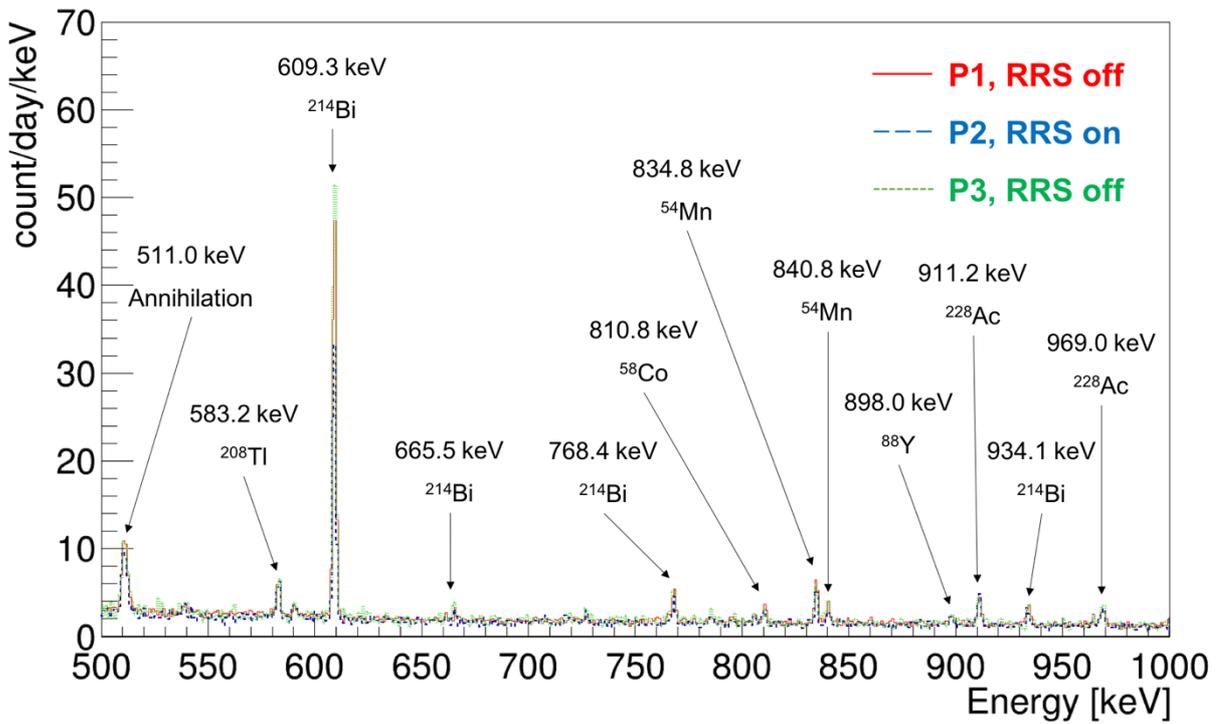

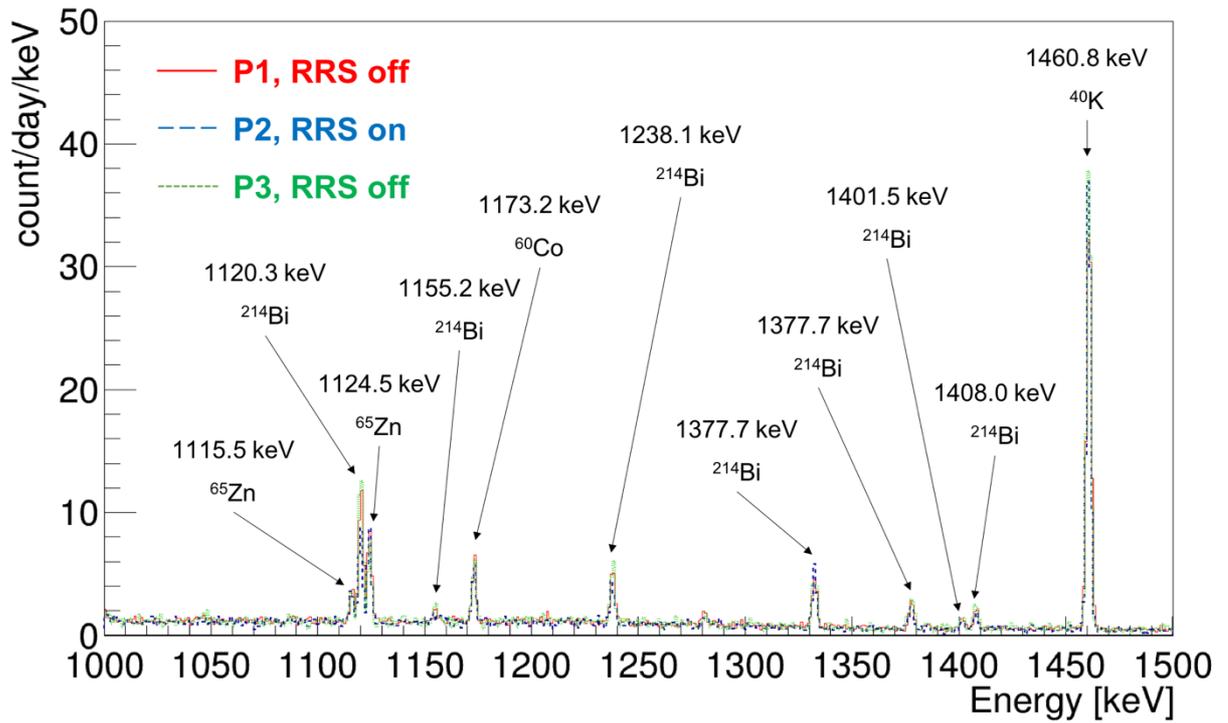
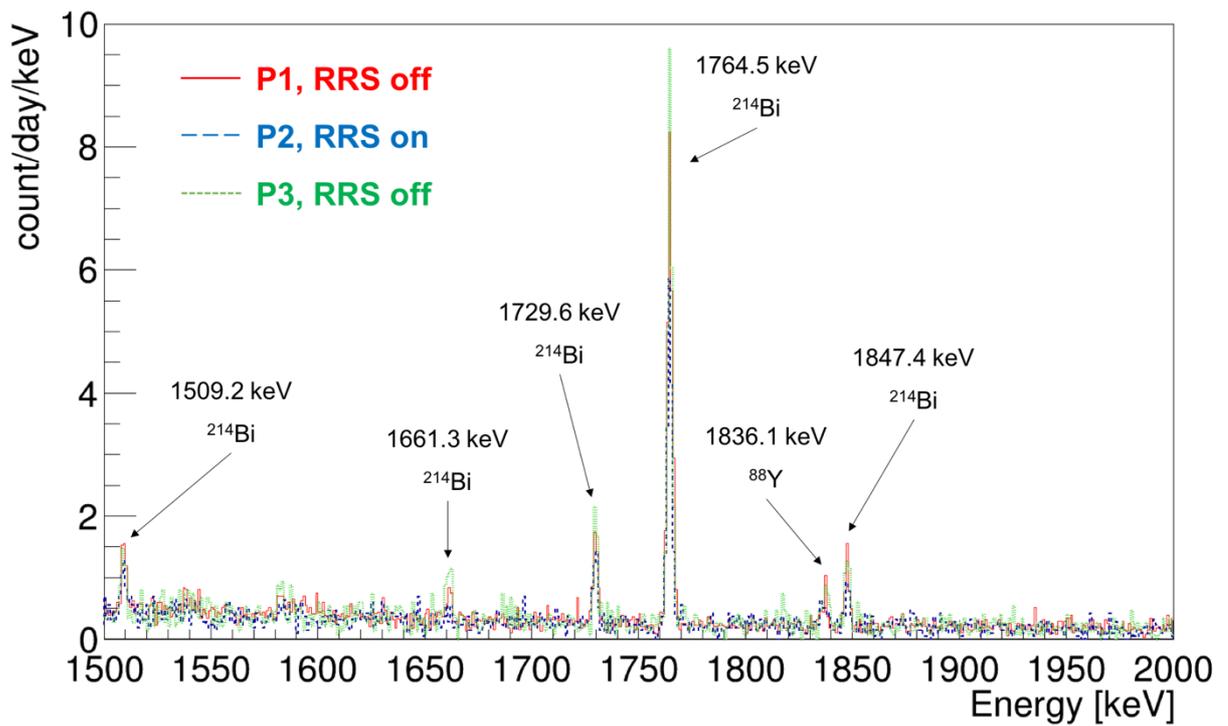

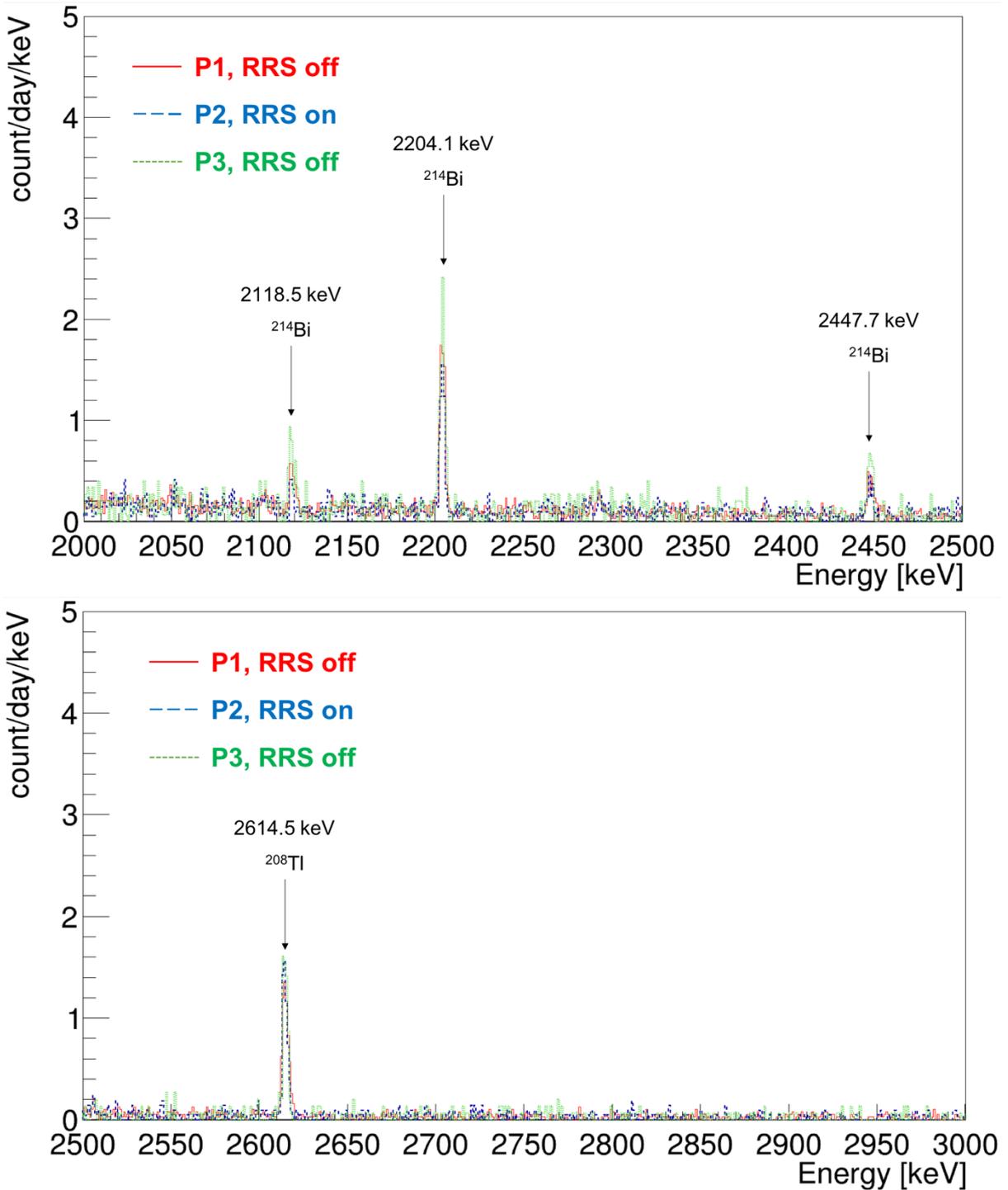

Fig. 3. Energy spectra from 40 keV to 3,000 keV divided into eight 500 keV regions. Red, blue, and green are data for counting periods P1, P2, and P3, respectively.

The count rate in the 609.3 keV peak was lower in P2 with RRS on than it was in P1 or P3 with the RRS off. On the other hand, count rates for the 583.2 keV peak from $^{208}$Tl, for the 898.0 keV peak from $^{88}$Y, and for the 1,460.8 keV peak from $^{40}$K were not affected significantly by the RRS operation. Comparisons of those peaks are shown in Fig. 4. Background data sets had the same tendency.

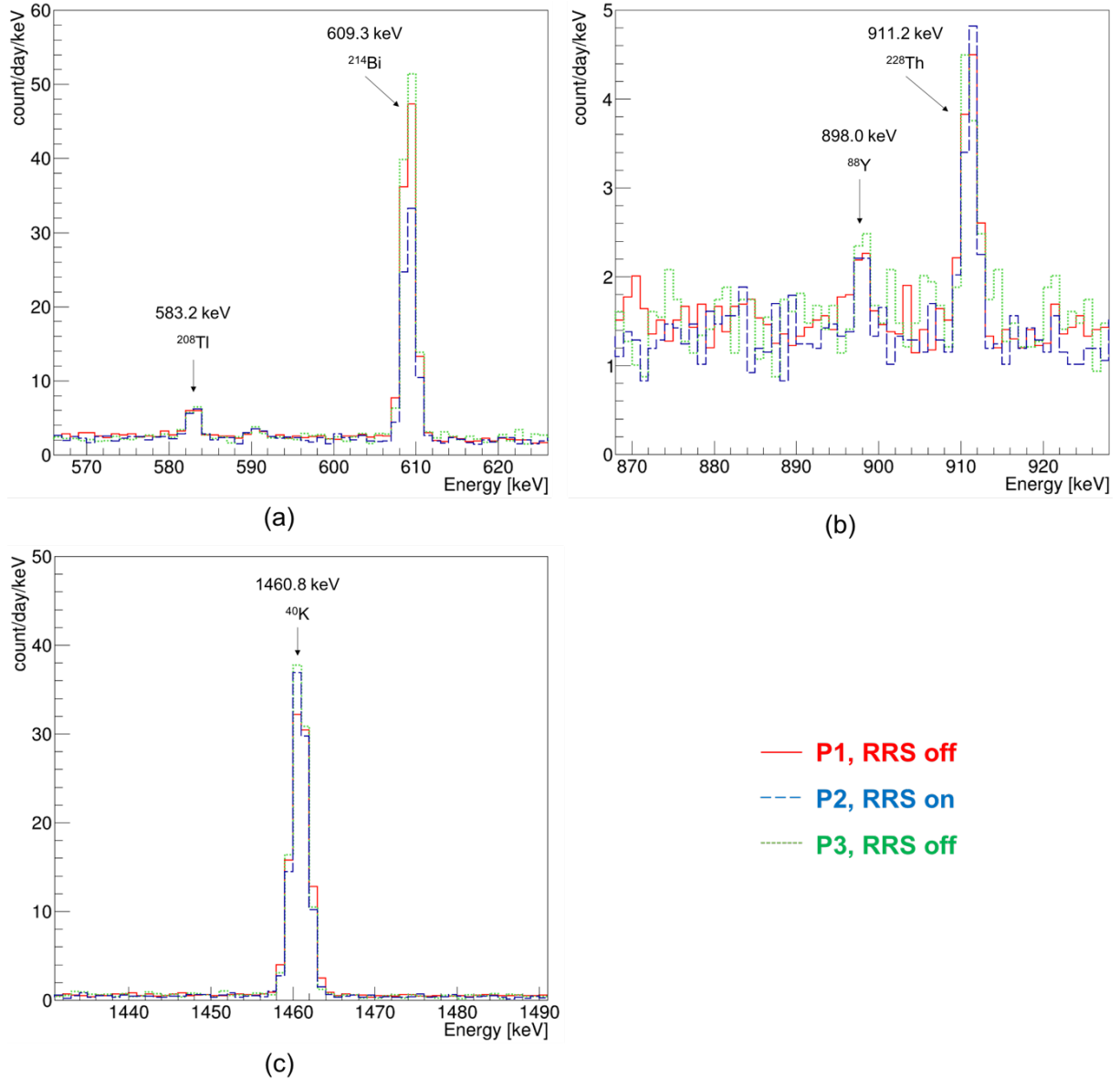

Fig. 4. Some prominent peaks from powder counting data. The (a) 583.2 keV, 609.3 keV, (b) 898.0 keV, and (c) 1,460.8 keV peaks and their sources are shown, comparing data from the counting periods P1 (red), P2 (blue), and P3 (green). P1 and P3 had the RRS off while P2 had the RRS operating.

## 5. Analysis

All radioactive-decay data used in the activity analysis were from the National Nuclear Data Center (NNDC) [31]. We calculated detection efficiencies using the GEANT4 Monte Carlo simulation program [32]. ROOT was used to facilitate count rate analysis of peaks in the energy spectra [33].

We do not assume full chain equilibrium for the full $^{232}$Th and $^{238}$U decay chains. Rather, we report activities of long-lived isotopes which support sub-chain decays. These isotopes are $^{228}$Ra and $^{228}$Th in the $^{232}$Th chain and $^{226}$Ra in the $^{238}$U chain. We do assume equilibrium within the sub-chains supported by these decays, which is guaranteed under most conditions.

$^{228}$Ra has a half-life of 6.7 yrs and the only observed decay is from $^{228}$Ac with a half-life of 6.1 min. This implies that any sample that has been chemically stable on the scale of less than an hour is in equilibrium and that the $^{228}$Ac activity represents the $^{228}$Ra sub-chain activity. The $^{226}$Ra sub-chain stabilizes in time scales on the order of tens of minutes associated with the $^{214}$Pb and $^{214}$Bi half-lives. Likewise, the $^{228}$Th sub-chain is stabilized with the 3.6 days half-life of $^{224}$Ra. For very freshly purified samples, it is possible to see breakage in this sub-chain, but any samples more than about a month old will be in equilibrium.

Because the CAGe is segmented, events can be classified based on the number of detector hits, i.e. the number of detectors with energy depositions above the signal threshold 50 ADC channels, corresponding to about 40 keV. For instance, the description "single-hit" means the event has only a single detector with enough energy deposit, and "double-hit" describes an event when two individual detectors each recorded energy deposits above the threshold [20].

Fig. 5 shows the number of events vs. the number of detector hits for all powder data from twelve detectors. Of all events, 91.59% were single-hit. However, a significant fraction, 7.73%, were double-hit events. Events hitting three or more detectors made up a negligible portion of the data set. For most activities, peaks were clearer and the sensitivities were larger for the single-hit spectrum than for the multiple-hit spectra. Single-hit analysis was performed for all radioisotopes. However, there were clear peaks from $^{228}$Th decay in the double-hit spectrum and these were analyzed in addition to the peaks in the single-hit spectrum.

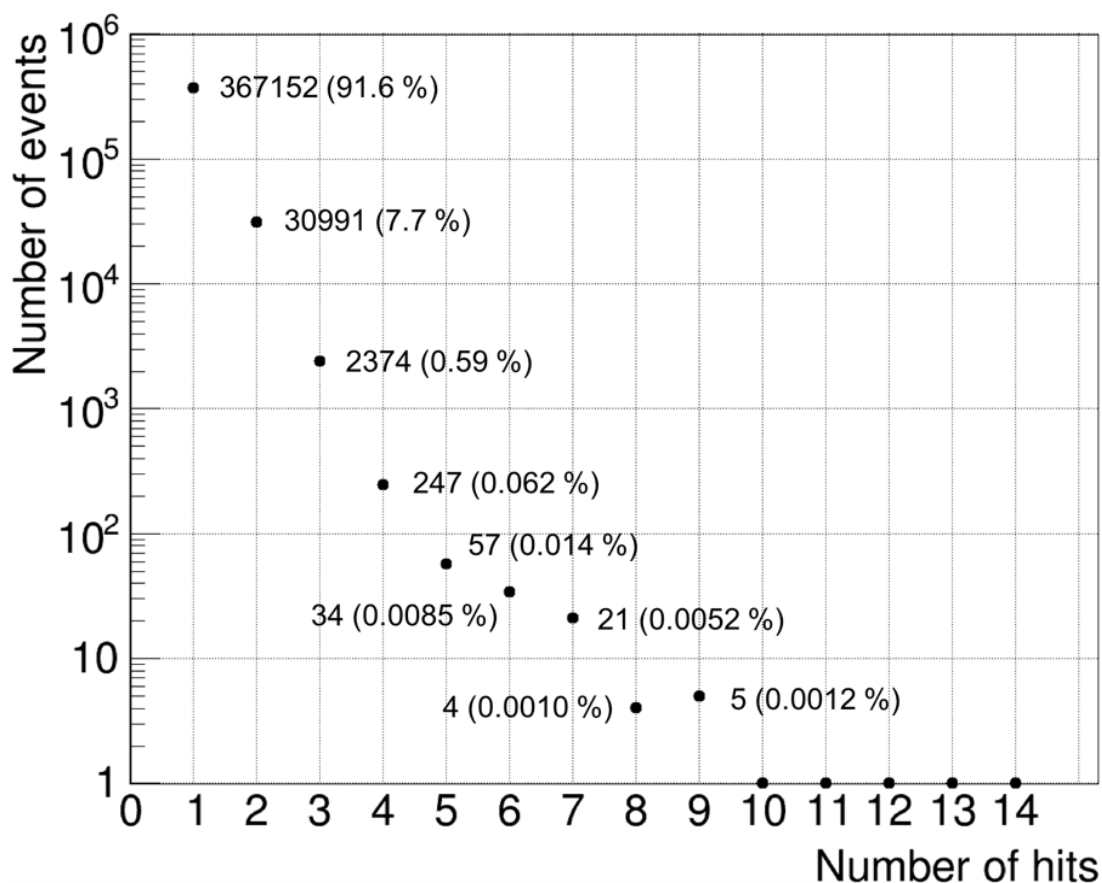

Fig. 5. The number of events vs. the number of detectors with signals above the threshold.

## 5.1 Single-hit Analysis

For $^{228}$Ac, $^{228}$Th, $^{88}$Y, and $^{40}$K which were not affected by the RRS operation, activities were analyzed in a single step using the combined sample data from P1, P2, and P3, and combined background data from B1 and B2. The spectra of single-hit events for powder data and background data are shown in Fig. 6.

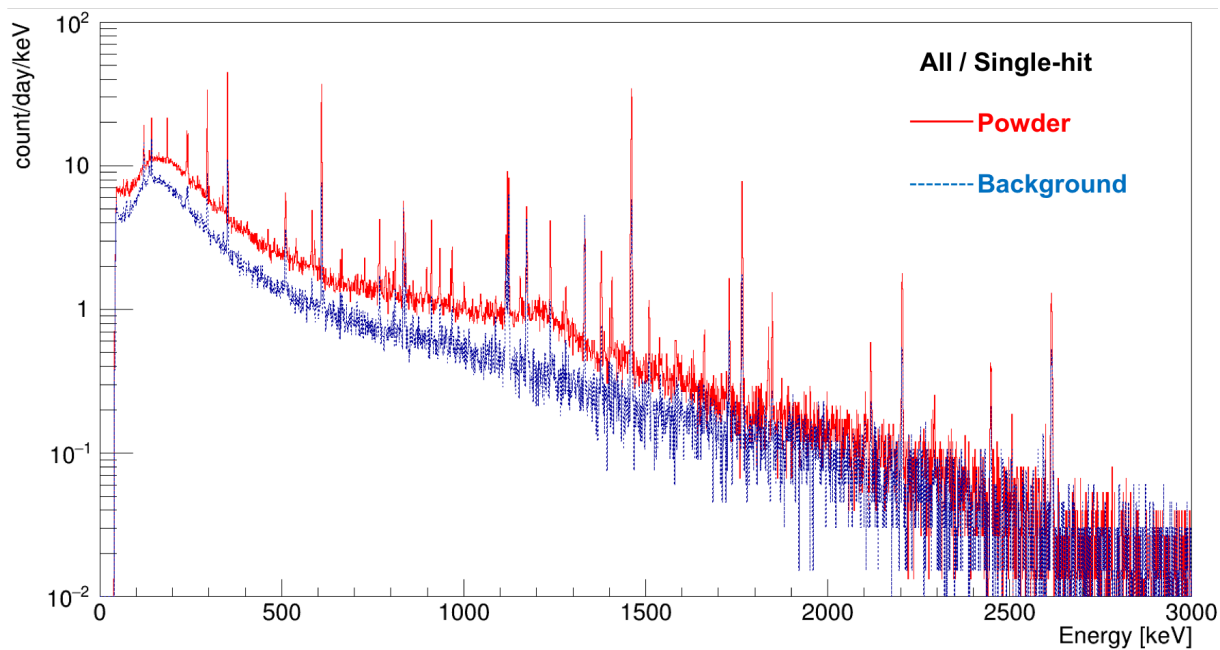

Fig. 6. Single-hit energy spectra for both the powder data and the background data. They were used for the analysis of $^{228}$Ac, $^{228}$Th, $^{88}$Y, and $^{40}$K.

The rates of the $^{226}$Ra-related backgrounds with the RRS on were different from the rates with it off (See Sec. 4). Therefore, the data from period B1 with RRS off was used for $^{226}$Ra background analysis of the combined P1 and P3 data sets, while the B2 data with RRS on was for $^{226}$Ra background analysis of the P2 data.

Fig. 7 compares powder and background spectra for both the RRS-on data and the RRS-off data. Since radon only impacts the $^{226}$Ra-related rates, this separation was only done for this analysis.

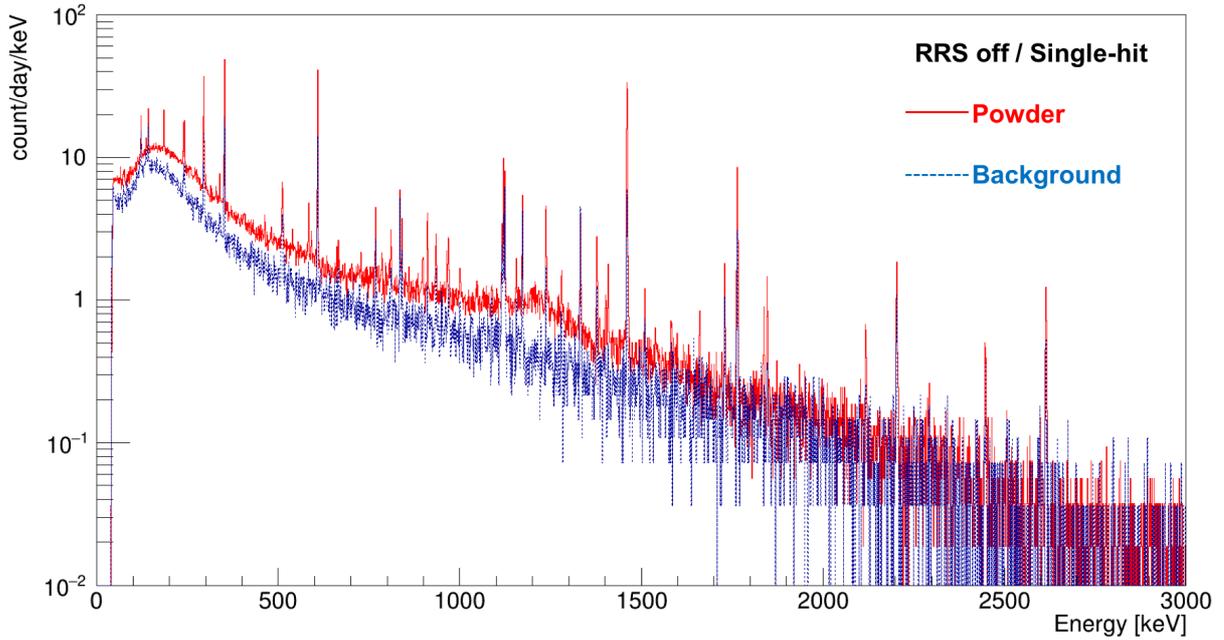

(a)

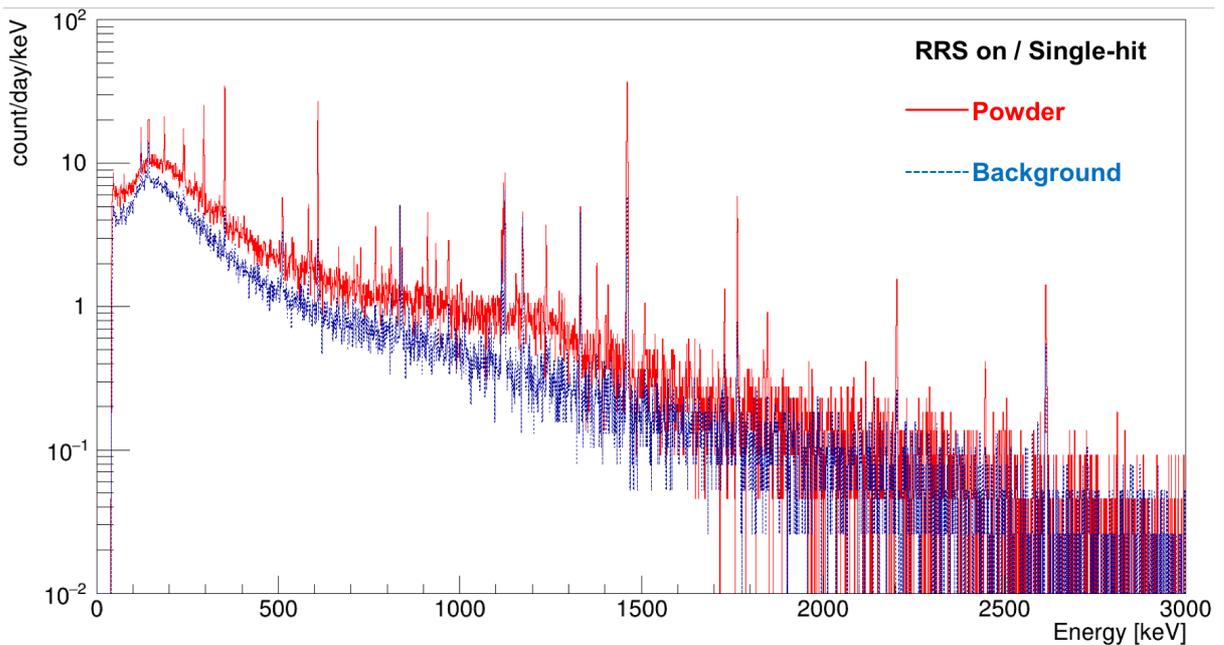

(b)

Fig. 7. (a) The energy spectrum with the RRS off for combined P1 and P3 data set analysis, and (b) the spectrum with the RRS on for P2 analysis.

## 5.2 Double-hit Analysis

The $\beta^-$ decay of $^{208}$Tl produces at least two coincident gammas for nearly 100% of decays. The 2,614.5 keV gamma is emitted in coincidence with at least one of a few other gamma emissions such as the 583.2 keV gamma. Similarly, $^{88}$Y usually emits two coincident gammas of 898.0 keV and 1,836.1 keV. By identifying coincidence events we can gain further specificity and background reduction, but at the expense of greatly reduced efficiency for these event classes.

As shown in Fig. 8, two-dimensional energy spectra of double-hit events are generated where the higher energy, $E_1$, is always assigned to the x-axis, and the lower energy, $E_2$ is assigned to the y-axis. Coincidence events were identified by requiring two-hit events with energy deposition in one detector equal to the full energy, $E_\gamma$, of one of the emitted gammas of a known coincidence pair. A loose cut around $E_\gamma$ is applied to either $E_1$ or $E_2$, separately. This cut is essentially aesthetic as the peak is ultimately extracted by fitting in this dimension, not by cutting. However, when the $E_\gamma$ cut is applied to $E_1$ ($E_2$) then a second cut is applied on the energy, $E_2$ ($E_1$), deposited in the second detector, simply to restrict it to be less than the full energy of any coincident gamma, thus eliminating some extraneous background events from the peak. In this way, selecting $E_\gamma$ from each of $E_1$ and $E_2$ creates vertical and horizontal bands in the two-dimensional spectra.

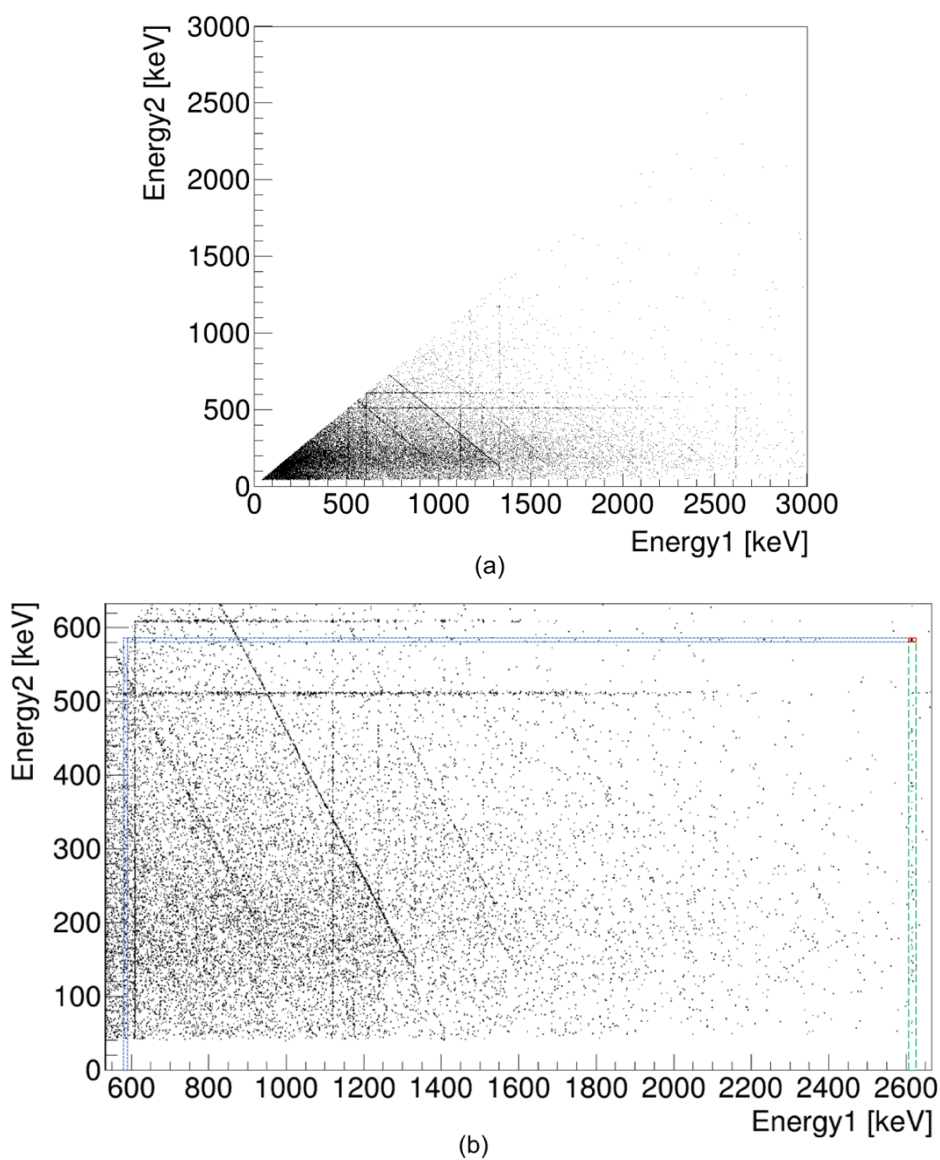

Fig. 8. [Color online] Two-dimensional energy distributions of double-hit events. The distributions are for (a) 40 keV to 3,000 keV in both detectors, and (b) zoomed in to see the $^{208}$Tl coincidence regions.

For example, the blue band in Fig. 8(b) includes events where a 583.2 keV gamma is fully deposited in one detector and a deposition less than 2,614.5 keV - $3\sigma_{2614.5\ keV}$ is found in the second detector. Similarly, the green band represents the deposition of the full energy of a 2,614.5 keV gamma along with a partial deposition in a second detector less than 583.2 keV - $3\sigma_{583.2\ keV}$. The two widths of $\sigma_{583.2\ keV}$ and $\sigma_{2614.5\ keV}$ here are from Gaussian fits of the 583.2 keV and the 2,614.5 keV peaks, respectively. Simultaneous full-energy deposition of both 583.2 keV and 2,614.5 keV gammas in separate detectors is also rare and was considered as a separate event class contained in the red box cut as shown in Fig. 8(b).

After applying the 2-dimensional box cuts, the resulting 1-dimensional $E_1$ and $E_2$ spectra were summed into a single spectrum. In this way, both the vertical and horizontal 583.2 keV bands in Fig. 8 were merged into a single peak for fitting. There were no statistically significant peaks for the 898.0 keV and 1,836.1 keV from $^{88}$Y in the double-hit spectrum as shown in Figs. 9(a) and 9(b). The two peaks at 583.2 keV and 2,614.5 keV were each fitted with a background component, thus removing any remaining Compton background.

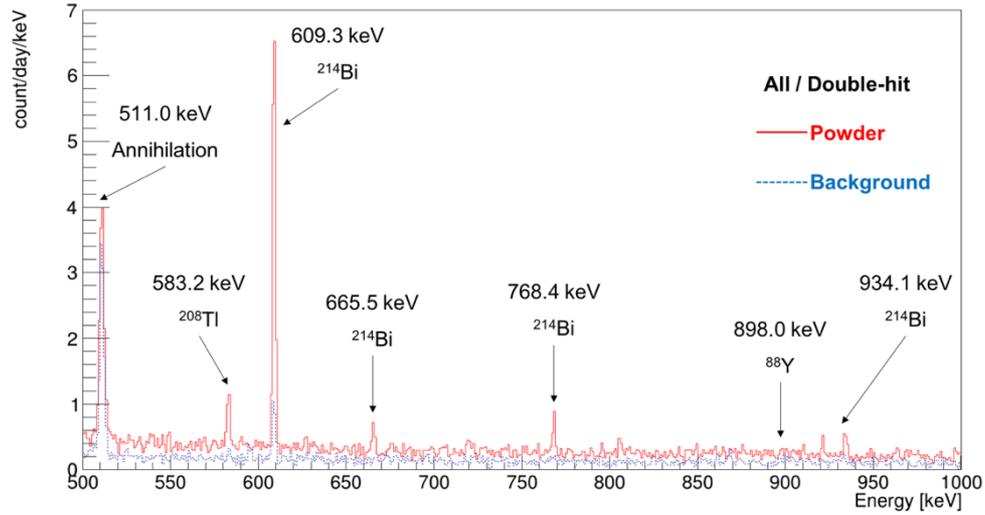

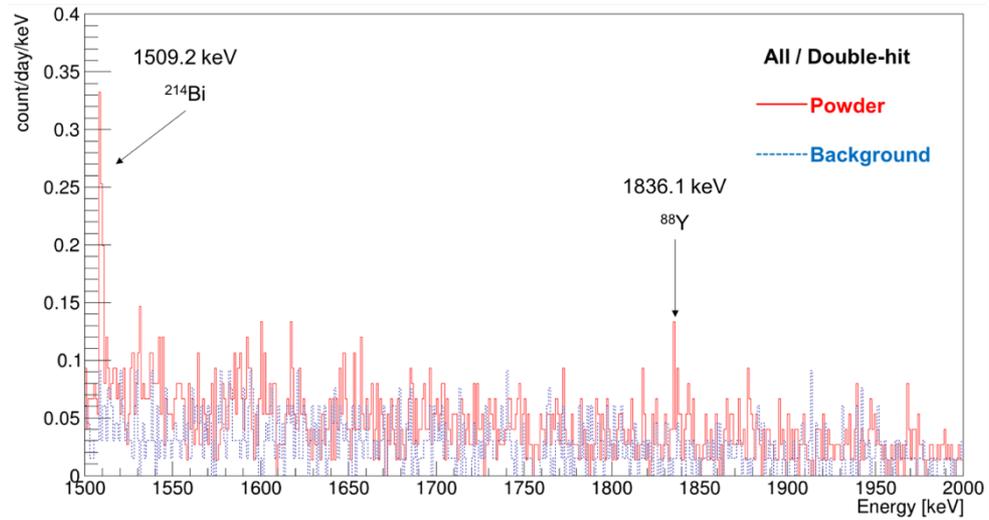

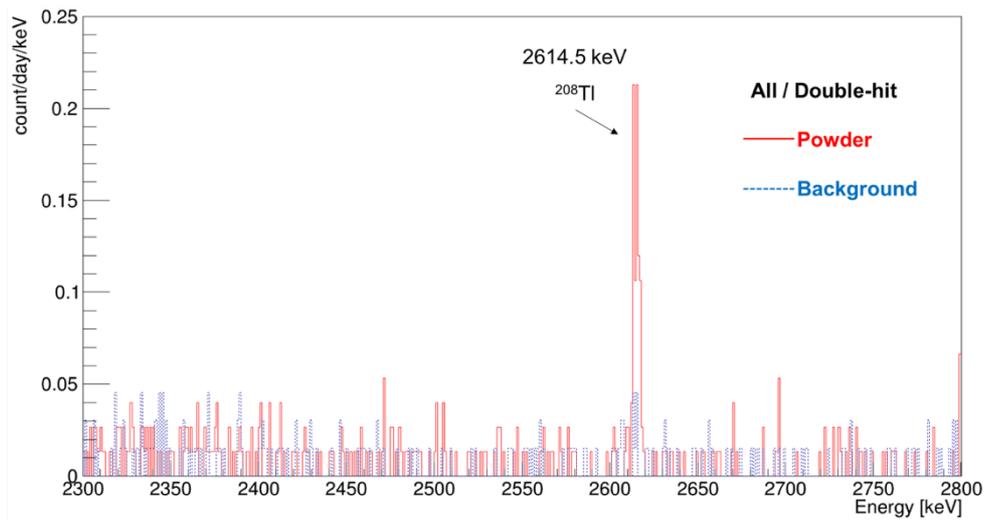

Fig. 9. One-dimensional spectra of double-hit events. Peaks of (a) 583.2 keV, 898.0 keV, (b) 1,836.1 keV, and (c) 2,614.5 keV from $^{208}$Tl and $^{88}$Y.

## 5.3 Background Activities from Labels

Information labels were attached to all the plastic bags which contained the $^{100\text{Enr}}\text{MoO}_3$ powder. Fourteen pieces of labels in total, which were attached to each of the selected fourteen powder bags, were kept on the bags during the measurement. Logistical constraints, along with the need to prevent contamination or sample loss, prevented removal of the labels before performing the activity measurement of this powder batch. The labels were installed together with the powder sample as shown in Figs. 1(b) and 1(c).

Ref. 21 describes how the labels influenced a similar powder sample measurement performed using a single-element detector. Separate activity measurements of the labels made using the CC1 detector were reported, and systematic corrections to the data analysis, with associated uncertainties, are described there. Signals contributed by contaminants in the labels were subtracted in the present study as well, following the same procedure. Six additional sets of label samples were assayed for the present work, and updated activities including the combined results from Ref. 21 and this work are reported in Table II.

**TABLE II**
Activities of various isotopes in the representative label samples were measured in units of mBq/label. Fourteen labels in total were installed together with the powder sample.

| Isotope | $^{228}\text{Ac}$ | $^{228}\text{Th}$ | $^{226}\text{Ra}$ | $^{40}\text{K}$ |
|---|---|---|---|---|
| Activity [mBq/label] | 0.074 ± 0.019 | 0.065 ± 0.011 | 2.23 ± 0.22 | 0.79 ± 0.13 |

## 5.4 Detection Efficiency and Simulation

To determine detection efficiencies, we simulated decay chains of $^{232}\text{Th}$ and $^{238}\text{U}$, as well as decays of $^{88}\text{Y}$ and $^{40}\text{K}$. The GEANT4 toolkit was used to perform the simulations. Geometries of the CAGe detector system and acrylic sample supporters were included in the simulation in detail, along with geometries of the powder sample and the labels as shown in Fig. 10. To subtract contributions from the labels, efficiencies were determined for sources distributed randomly, with a uniform probability distribution, within the powder sample and separately for sources distributed in the labels. We note that the uniform probability distribution imposes an assumption on the efficiency analysis that all powder bags had equal and well distributed concentrations of interest.

As was done in Ref. 21, uncertainty in the precise sample position was considered by varying the sample position within reasonable bounds in the Monte Carlo and finding the efficiencies for the different configurations. From the resulting differences, we estimate a systematic error in the efficiencies of about 5%.

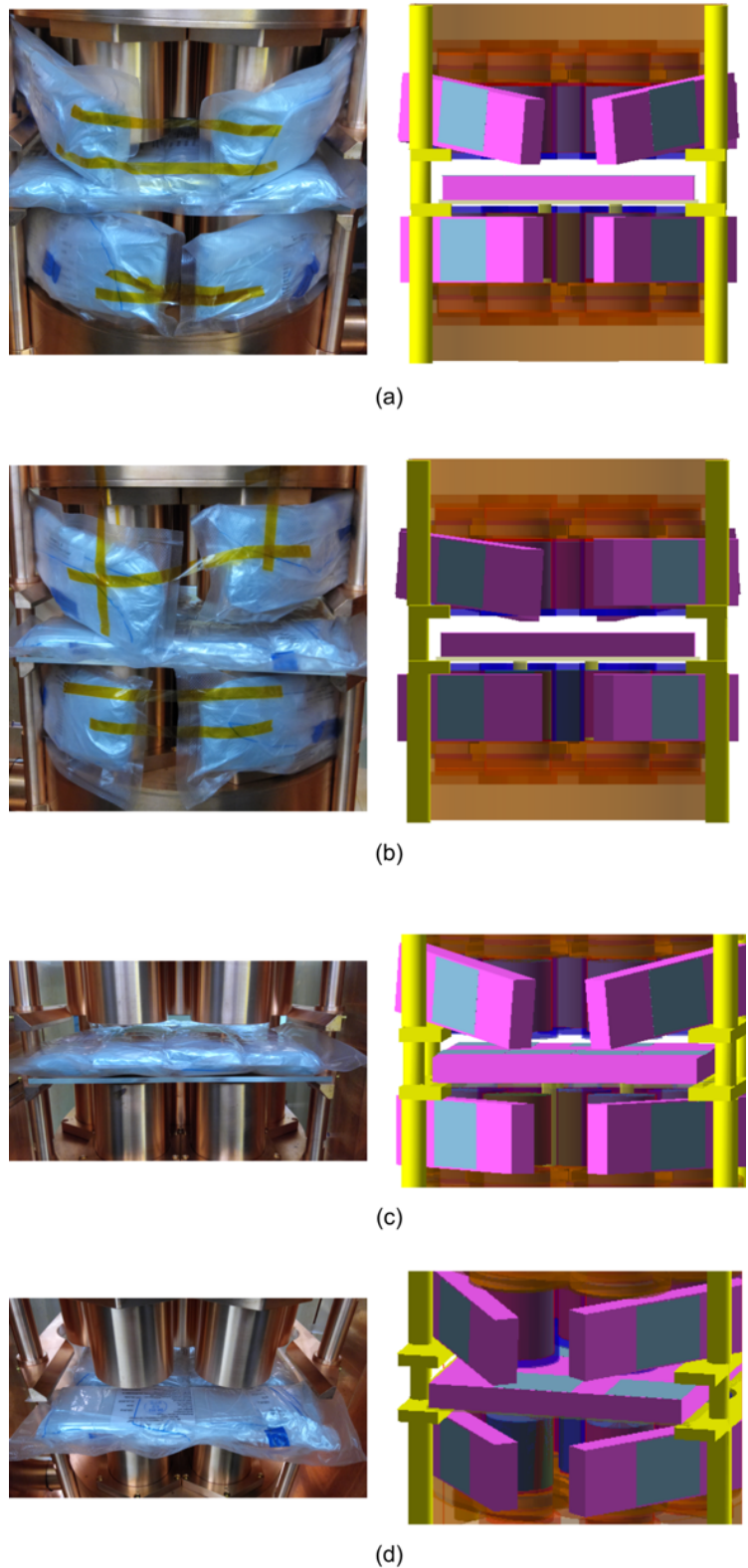

Fig. 10. Detailed construction of the CAGe detection system, the powder sample, and labels are shown in photos (left column) and rendered simulation geometries (right column). In the simulation depictions, the powder sample is in magenta, and the labels are in gray. The detector is shown from two opposite sides (a) and (b). The labels on the powder bags between the array elements can be seen from both sides in (c) and (d).

Detection efficiencies were calculated in the same way as described in Ref. 21. Detection efficiency, $\varepsilon$, of a gamma with energy $E_i$ is calculated as $\varepsilon = \frac{N_d}{N_i}$, where $N_d$ is the number of gammas detected in the full-energy peak at $E_i$, and $N_i$ is the number of gammas generated with energy $E_i$.

For activity analysis, we considered gammas emitted from high-intensity decays of $^{228}$Ac, $^{88}$Y, and $^{40}$K in addition to the decay chains of $^{228}$Th and $^{226}$Ra. Detection efficiencies, $\varepsilon_P$ and $\varepsilon_L$, for gamma transitions in the powder sample and the labels, respectively, are listed in Table III. No label contributions were expected, observed, or subtracted for the $^{88}$Y peaks, and the detection efficiencies of the corresponding 898.0 keV and 1,836.1 keV gammas from the label were not noted.

**TABLE III**
Radioisotopes and gamma energies used for $^{228}$Ac, $^{228}$Th, $^{226}$Ra, $^{88}$Y, and $^{40}$K activity analyses are categorized by multiple-hit criteria. Detection efficiencies, $\varepsilon_P$ and $\varepsilon_L$, of the powder sample and the labels, respectively, are listed for each gamma.

| Multiplicity | Supporting isotope | Decaying isotope | Energy [keV] | Detection efficiency (%) | |
|---|---|---|---|---|---|
| | | | | $\varepsilon_P$ | $\varepsilon_L$ |
| Single-hit | $^{228}$Ra | $^{228}$Ac | 911.2 | 2.76 | 2.89 |
| | | | 969.0 | 2.78 | 2.77 |
| Single-hit | $^{228}$Th | $^{208}$Tl | 238.6 | 4.46 | 4.50 |
| | | | 583.2 | 2.78 | 2.71 |
| | | | 2,614.5 | 1.27 | 1.30 |
| Double-hit | $^{228}$Th | $^{208}$Tl | 583.2 | 0.61 | 0.71 |
| | | | 2,614.5 | 0.28 | 0.34 |
| Single-hit | $^{226}$Ra | $^{214}$Pb | 352.0 | 4.52 | 4.65 |
| | | $^{214}$Bi | 609.3 | 2.83 | 2.84 |
| | | | 1,120.3 | 2.34 | 2.37 |
| | | | 1,764.5 | 2.24 | 2.24 |
| Single-hit | $^{88}$Y | $^{88}$Y | 898.0 | 2.42 | N/A * |
| | | | 1,836.1 | 1.73 | N/A |
| Single-hit | $^{40}$K | $^{40}$K | 1,460.8 | 2.45 | 2.47 |

\* Not applied N/A

A potential uncertainty in the simulation arises from inactive surface regions of the germanium crystals, regions which are called dead layers [17, 34]. These act as shielding layers and also reduce the detection volume of germanium. We performed detection efficiency calibration to ascertain the thickness of these dead layers by assaying a source, containing ten different radioisotopes, consecutively positioned at the end of each detector element of the CAGe.

Overall, a 12% systematic error in the detection efficiency was found, including dead-layer uncertainty and geometrical uncertainties in the detector, the powder sample, and the labels geometry.

### 5.5 Count and Activity

The activity $A_P$ of an isotope of interest in the powder sample, determined from a single gamma peak in its decay chain, is given in Ref. 21 as

$$A_P = \frac{\frac{C_O}{T_P} - \frac{C_L}{T_P} - \frac{C_B}{T_B}}{\varepsilon_P \times M_P \times B.R. \times G.I.},$$

where $C_O$ is the number of counts observed in the peak of interest during the powder measurement time $T_P$, $C_L$ is the estimated number of counts contributed to the peak during the same period by the labels, $C_B$ is the observed number of counts in the peak during the background measurement time $T_B$, $M_P$ is the mass of the powder sample, $B.R.$ is branching ratio of the gamma-emitting decay relative to the isotope of interest, and $G.I.$ is the gamma intensity from the gamma-emitting decay. As described in Sec. 5.4, $\varepsilon_P$ is the detection efficiency of the full-energy peak for the gamma of interest with the source decays distributed within the powder sample.

The estimated count rate, $\frac{C_L}{T_P}$, from the labels only is obtained by

$$\frac{C_L}{T_P} = A_L \times \varepsilon_L \times N_L \times B.R. \times G.I.,$$

where $A_L$ is the activity of the labels in Table II, and $N_L$ is the total number of the labels assayed with the powder sample in the CAGe. Again, $\varepsilon_L$ is the peak detection efficiency corresponding to decays distributed in the labels.

The count rates, $\frac{C_O}{T_P}$, $\frac{C_L}{T_P}$, and $\frac{C_B}{T_B}$, for each gamma from $^{228}$Ac, $^{228}$Th, $^{226}$Ra, $^{88}$Y, and $^{40}$K decays are listed in Table IV.

## TABLE IV

Radioisotopes and gamma energies used for $^{228}$Ac, $^{228}$Th, $^{226}$Ra, $^{88}$Y, and $^{40}$K activity analysis categorized by RRS-operation and multiplicity are summarized. Observed total count rate, $\frac{C_O}{T_P}$, from the powder sample, the labels, and the CAGe detector system are listed for each gamma as well as estimated count rate $\frac{C_L}{T_P}$ from the labels alone, and count rate $\frac{C_B}{T_B}$ from the background data. The decaying isotope associated with the gamma emission is listed. As most of these have very short half-lives, the long-lived isotopes which directly support the observed decays are also listed. For comparison, counts are normalized to acquisition times $T_{P,\,All}$ (1,803.3 hrs), $T_{P,\,RRS\,off}$ (1,280.3 hrs), $T_{P,\,RRS\,on}$ (523.1 hrs), $T_{B,\,All}$ (1,582.5 hrs), $T_{B,\,RRS\,off}$ (663.1 hrs), and $T_{B,\,RRS\,on}$ (919.4 hrs).

| Data | Multiplicity | Supporting isotope | Decaying isotope | Energy [keV] | Count rate [× $10^{-2}$ counts/hour] | | |
|---|---|---|---|---|---|---|---|
| | | | | | $\frac{C_O}{T_P}$ | $\frac{C_L}{T_P}$ | $\frac{C_B}{T_B}$ |
| All | Single-hit | $^{228}$Ra | $^{228}$Ac | 911.2 | 30.6 ± 1.9 | 2.78 ± 0.70 | 5.8 ± 1.1 |
| | | | | 969.0 | 20.0 ± 1.6 | 1.63 ± 0.41 | 5.1 ± 1.1 |
| All | Single-hit | $^{228}$Th | $^{208}$Tl | 238.6 | 69.0 ± 3.6 | 6.5 ± 1.1 | 21.4 ± 4.6 |
| | | | | 583.2 | 26.8 ± 1.7 | 2.72 ± 0.46 | 6.0 ± 1.5 |
| | | | | 2,614.5 | 20.2 ± 1.1 | 1.54 ± 0.26 | 7.39 ± 0.74 |
| All | Double-hit | $^{228}$Th | $^{208}$Tl | 583.2 | 7.13 ± 0.98 | 0.71 ± 0.12 | 1.30 ± 0.50 |
| | | | | 2,614.5 | 3.30 ± 0.45 | 0.397 ± 0.066 | 0.20 ± 0.18 |
| RRS off | Single-hit | $^{226}$Ra | $^{214}$Pb | 352.0 | 416.0 ± 6.3 | 186 ± 19 | 153.1 ± 5.7 |
| | | | $^{214}$Bi | 609.3 | 343.0 ± 5.5 | 146 ± 15 | 127.9 ± 4.9 |
| | | | | 1,120.3 | 94.6 ± 3.2 | 39.8 ± 4.0 | 34.2 ± 2.8 |
| | | | | 1,764.5 | 97.0 ± 2.9 | 38.5 ± 3.9 | 31.3 ± 2.4 |
| RRS on | Single-hit | $^{226}$Ra | $^{214}$Pb | 352.0 | 278.3 ± 8.2 | 186 ± 19 | 22.3 ± 2.8 |
| | | | $^{214}$Bi | 609.3 | 220.2 ± 7.1 | 146 ± 15 | 24.1 ± 2.3 |
| | | | | 1,120.3 | 58.7 ± 4.0 | 39.8 ± 4.0 | 10.1 ± 1.5 |
| | | | | 1,764.5 | 64.5 ± 3.7 | 38.5 ± 3.9 | 7.6 ± 1.2 |
| All | Single-hit | $^{88}$Y | $^{88}$Y | 898.0 | 8.5 ± 1.4 | N.M. * | N.M. |
| | | | | 1,836.1 | 6.7 ± 1.2 | N.M. | N.M. |
| All | Single-hit | $^{40}$K | $^{40}$K | 1,460.8 | 394.9 ± 4.8 | 10.5 ± 1.8 | 59.7 ± 2.0 |

\* Not-measured N.M.

## 6. Result and Conclusion

As described in Sec. 5, $^{228}$Th analysis was performed separately for single-hit and double-hit events. The results were combined by using the weighted mean with statistical errors only, applying systematic errors to the final result. Likewise, results for $^{226}$Ra obtained from data with the RRS on and off were also combined by using the weighted mean. Final results were obtained by using the error propagation of statistical and systematic uncertainties and are summarized in Table V.

**TABLE V**
Activities of $^{228}$Ac, $^{228}$Th, $^{226}$Ra, $^{88}$Y, and $^{40}$K in the powder sample. The results are based on the assumption of uniform distribution of contaminants within the powder.

| Isotope | $^{228}$Ac | $^{228}$Th | $^{226}$Ra | $^{88}$Y | $^{40}$K |
|---|---|---|---|---|---|
| Activity [mBq/kg] | $0.88 \pm 0.12$ | $0.669 \pm 0.087$ | $1.50 \pm 0.23$ | $0.101 \pm 0.016$ | $36.0 \pm 4.1$ |

The activities of radioisotopes in the powder sample, powder which will be used in crystal growing for the AMoRE-II double-beta decay search, were measured using the CAGe detector system. These activities must be quantified and controlled in order to constrain the backgrounds from radioactivity in the final crystals for AMoRE-II.

The activity of the cosmogenic isotope $^{88}$Y depends strongly on the history of the material. This batch was delivered by a few air-flights, briefly elevating its cosmogenic exposure greatly, and was then stored for 259 days underground at the Y2L before counting. In comparison, the $^{88}$Y half-life is 106.6 days [11].

As mentioned in Sec. 1, the previous measurement of Ref. 21, using the single-element CC1 detector, produced an upper limit for the $^{228}$Th activity. Improved sensitivity of the CAGe relative to the CC1 detector helped to achieve sensitivities at the level of a few hundred μBq/kg. $^{228}$Th activity was observed with over 3σ significance, representing an important result for background control in the AMoRE-II experiment.

**Acknowledgments**

This work was supported by the Institute for Basic Science (IBS) funded by the Ministry of Science and ICT, Korea (Grant id: IBS-R016-D1).

**References**

[1] S. Rahaman, et al., Q values of the $^{76}$Ge and $^{100}$Mo double-beta decays, Phys. Lett. B, 662 (2008) 111.

[2] A. S. Barabash, V. B. Brudanin, Investigation of Double-Beta Decay with the NEMO-3 Detector, Phys. Atom. Nucl., 74 (2011) 312.

[3] H. K. Park, et al., (AMoRE collaboration), The AMoRE: Search for Neutrinoless Double Beta Decay in $^{100}$Mo, Nucl. Part. Phys. Proc., 273-275 (2016) 2630.

[4] V. Alenkov, et al., First results from the AMoRE-Pilot neutrinoless double beta decay experiment, Eur. Phys. J. C, 79 (2019) 791.

[5] V. Alenkov, et al., Technical Design Report for the AMoRE 0νββ Decay Search Experiment, arXiv:1512.05957 [physics.ins-det].

[6] G. B. Kim, et al., A CaMoO4 Crystal Low Temperature Detector for the AMoRE Neutrinoless Double Beta Decay Search, Adv. High. Energy Phys., 2015 Article ID 817530 (2015).

[7] M. J. Martin, Nuclear Data Sheets for A = 208, Nuclear Data Sheets, 108 (2007) 1583.

[8] M. Shamsuzzoha Basunia, Nuclear Data Sheets for A = 210, Nuclear Data Sheets, 121 (2014) 561.

[9] S. -C. Wu, Nuclear Data Sheets for A = 214, Nuclear Data Sheets, 110 (2009) 681.

[10] D. Cherniak, Development of cryogenic low background detector based on enriched zinc molybdate crystal scintillators to search for neutrinoless double beta decay of $^{100}$Mo, arXiv:1507.04591 [physics.ins-det].

[11] E. A. McCutchan, A. A. Sonzogni, Nuclear Data Sheets for A = 88, Nuclear Data Sheets, 115 (2014) 135.

[12] JUN CHEN, Nuclear Data Sheets for A = 40, Nuclear Data Sheets, 140 (2017) 1.

[13] J. H. So, et al., Scintillation Properties and Internal Background Study of $^{40}$Ca$^{100}$MoO$_4$ Crystal Scintillators for Neutrino-Less Double Beta Decay Search, IEEE Trans. Nucl. Sci., 59 (2012) 2214.

[14] J. Y. Lee, et al., A Study of Radioactive Contamination of $^{40}$Ca$^{100}$MoO$_4$ Crystals for the AMoRE Experiment, IEEE Trans. Nucl. Sci., 63 (2016) 543.

[15] J. Y. Lee, et al., A Study of $^{48depl}$Ca$^{100}$MoO$_4$ Scintillation Crystals for the AMoRE-I Experiment, IEEE Trans. Nucl. Sci., 65 (2018) 2041.

[16] M. H. Lee, et al., An ultra-low radioactivity measurement HPGe facility at the Center for Underground Physics, in: Proc. The 39th International Conference on High Energy Physics (ICHEP2018), Seoul, Korea, 2018, 363.

[17] E. K. Lee, et al., Measurements of detector material samples with two HPGe detectors at the YangYang underground Laboratory, in: Proc. The 39th International Conference of High Energy Physics (ICHEP2018), Seoul, Korea, 2018, 809.

[18] A. Luqman, et al., Simulations of background sources in AMoRE-I experiment, Nucl. Instrum. Meth. A, 855 (2017) 140.

[19] O. Gileva, et al., Investigation of the molybdenum oxide purification for the AMoRE experiment, J Radioanal Nucl Chem, 314 (2017) 1695.

[20] G. W. Kim, Rare Decay Experiments for $^{180m}$Ta and $^{208}$Pb* Using HPGe Detectors, Ph. D. Dissertation, Ewha Womans University, 2019.

[21] S. Y. Park, et al., Measurement of the background activities of a $^{100}$Mo-enriched powder sample for an AMoRE crystal material by using a single high purity germanium detector, J. Korean Phys. Soc. 76 (2020) 1060.

[22] E. Sala, et al., Development of an underground HPGe array facility for ultra low radioactivity measurements, AIP Conference Proceedings, 1672, (2015) 120001.

[23] E. Sala, et al., Development of an underground low background instrument for high sensitivity measurements, J. Phys. Conf. Ser., 718 (2016) 062050.

[24] G. W. Kim, et al., Simulation Study for the Half-Life Measurement of $^{180m}$Ta Using HPGe Detectors, J. Korean Phys. Soc., 75 (2019) 32.

[25] D. S. Leonard, et al., Development of an array of HPGe detectors with 980% relative efficiency, arXiv:2009.00483v1 [physics.ins-det]

[26] M. H. Lee, Radioassay and Purification for Experiments at Y2L and Yemilab in Korea, J. Phys.: Conf. Ser., 1468 (2020) 012249.

[27] RAD7 radon detector, available at https://durridge.com/products/rad7-radon-detector/ (accessed on June 2nd, 2020).

[28] K. Szymańska, et al., Resolution, efficiency and stability of HPGe detector operating in a magnetic field at various gamma-ray energies, Nucl. Instrum. Meth. A, 592 (2008) 486.

[29] I. Hossain, N. Sharip, K. K. Viswanathan, Efficiency and resolution of HPGe and NaI(Tl) detectors using gamma-ray spectroscopy, Sci. Res. Essays, 7(1) (2012) 86.

[30] E. Browne, J. K. Tuli, Nuclear Data Sheets for A = 60, Nuclear Data Sheets, 114 (2013) 1849.

[31] National Nuclear Data Center, available at https://www.nndc.bnl.gov/nudat2/ (accessed on June 2nd, 2020).

[32] S. Agostinelli, et al., Geant4—a simulation toolkit, Nucl. Instrum. Meth. A, 506 (2003) 250.

[33] Rene Brun and Fons Rademakers, ROOT - An Object Oriented Data Analysis Framework, Proceedings AIHENP'96 Workshop, Lausanne, Sep. 1996, Nucl. Inst. & Meth. in Phys. Res. A 389 (1997) 81.

[34] Glenn F. Knoll, Radiation detection and measurement, 3rd ed., John Wiley & Sons, New York, 2009.